\documentclass[acmsmall]{acmart}
\AtBeginDocument{%
  \providecommand\BibTeX{{%
    \normalfont B\kern-0.5em{\scshape i\kern-0.25em b}\kern-0.8em\TeX}}}

\setcopyright{cc}
\setcctype{by}
\acmJournal{PACMHCI}
\acmYear{2025} \acmVolume{9} 
\acmNumber{7} 
\acmArticle{CSCW389}
\acmMonth{11} 
\acmPrice{}\acmDOI{10.1145/3757570}

\usepackage{soul}
\soulregister\cite7
\soulregister\citet7
\soulregister\ref7
\setstcolor{red}

\begin{document}

\title{Conflict in Community-Based Design: A Case Study of a  Relationship Breakdown}

\author{Alekhya Gandu}
\affiliation{%
  \institution{San Francisco State University}
  \country{San Francisco, CA, USA}
}
\email{alekhyahello@gmail.com}

\author{Aakash Gautam}
\affiliation{%
  \institution{University of Pittsburgh}
  \country{Pittsburgh, PA, USA}
}
\email{aakash@pitt.edu}

%\renewcommand{\shortauthors}{XYZ, et al.}

\begin{abstract}

Community-based design efforts rightly seek to reduce the power differences between researchers and community participants by aligning with community values and furthering their priorities. However, what should designers do when key community members' practices seem to enact an oppressive and harmful structure? We reflect on our two-year-long engagement with a non-profit organization in southern India that supports women subjected to domestic abuse or facing mental health crises.
We highlight the organizational gaps in knowledge management and transfer, which became an avenue for our design intervention. During design, we encountered practices that upheld caste hierarchies. These practices were expected to be incorporated into our technology. Anticipating harms to indirect stakeholders, we resisted this incorporation. 
It led to a breakdown in our relationship with the partner organization.
Reflecting on this experience, we outline pluralistic pathways that community-based designers might inhabit when navigating value conflicts. These include making space for reflection before and during engagements, strategically repositioning through role reframing or appreciative inquiry, and exiting the engagement if necessary.

\end{abstract}

\begin{CCSXML}
<ccs2012>
   %     <concept_id>10003120.10003121.10003122</concept_id>
   %     <concept_desc>Human-centered computing~HCI design and evaluation methods</concept_desc>
   %     <concept_significance>300</concept_significance>
   <concept>
       <concept_id>10003120.10003121.10011748</concept_id>
       <concept_desc>Human-centered computing~Empirical studies in HCI</concept_desc>
       <concept_significance>500</concept_significance>
   </concept>
 </ccs2012>
\end{CCSXML}

\ccsdesc[500]{Human-centered computing~Empirical studies in HCI}
% \ccsdesc[300]{Human-centered computing~HCI design and evaluation methods}

\keywords{community-based, participatory, relations, negotiations, caste, confrontation, ethics, responsibility, practice}

\received{October 2024}
\received[revised]{April 2025}
\received[accepted]{August 2025}

\maketitle

\section{Introduction}

Community-based design is now a widely practiced approach within CSCW and HCI \cite{liang2021embracing, fox2014community, bagalkot2022embodied, brown2019some}.  
It has enabled our field to eschew techno-deterministic priorities and center research efforts that align with the community's needs and values.
The justification for technology design is centered around the community, as it should be.  
This alignment with community values and prioritizing local needs also helps reduce power imbalances between researchers and community participants \cite{dearden2018minimum, bratteteig2012disentangling, hussain2012participatory}.
Scholarship within participatory design, HCI4D, and ICTD has been advocating ways to ensure that researchers' efforts are aligned with the community's goals and values \cite{bjorgvinsson2012agonistic, harrington2019deconstructing, lu2023participatory}.
Our research approach deeply appreciates this value, and we seek to learn and grow in this direction.

While a community-based approach is crucial in addressing concerns of academic extraction and exploitation, it presents a complex ethical challenge: What do we do when key community members'  values seem to adhere to an oppressive structure?
In this paper, we reflect on our work with Seva\footnote{Pseudonym. It means ``service'' in the local language, which is a value central to all the staff members in the organization.}, a non-profit organization in Southern India, dedicated to assisting women subjected to domestic abuse and mental health crises. 
Our research team is based in the United States. We conducted the study remotely. It was the first step in a planned long-term collaboration.
The initial inquiry revealed significant organizational gaps in knowledge management and transfer, which subsequently became the focus of our design intervention. 
To this end, we developed a web application to address the organizational challenges, focusing on improving information management and knowledge sharing among staff members. 
The application included features for client onboarding, documentation of services, and generation of reports. 
However, as we iterated on the design, examining concrete aspects of the organization's practices, we noted practices that enacted caste discrimination within the organization. This practice was expected to be incorporated into our design. 
We sought to engage in resistance, hoping to create space for negotiation and action toward shared visions.
We failed. 
The collaboration broke down.

We provide a descriptive reflection of our experience.
We situate this as a case study, highlighting four elements that we believe are of interest to the CSCW community.
First, we detail findings from our initial inquiry, which revealed gaps in knowledge management and discretionary individual actions.
Second, we briefly describe our design artifact, which was developed to support the organization in managing clients and organizing their collective knowledge.
Third, we recount the tensions that arose when staff members requested that we include a free-text field to record clients’ caste. We resisted this request, as we believed it could harm indirect stakeholders (their clients) by reinforcing caste-based hierarchies and exposing them to further marginalization. 
Our attempt to arrive at a compromise failed, leading to a breakdown in the collaboration.
Given how subtly caste discrimination is enacted, we relied on what we perceived rather than what was explicitly stated.
We justify our decisions based on retrospective inspection of earlier interviews and observations.
Our fourth contribution focuses on outlining strategies for designers handling conflicts in community engagements. These include creating opportunities for reflection before and during engagements, strategically distancing through role reframing or appreciative inquiry, and knowing when to exit the engagement if needed.

% We use “value conflict” to include cases where one party’s values—while harmful or exclusionary in effect—may not be explicitly recognized as such by the actors involved. Rather than pre-categorize such values as illegitimate, we examine how embedded practices—like caste-based identifiers—may be normalized by dominant groups and yet contradict the liberatory goals of design. Thus, we treat these as value conflicts with structural implications.
% While we, the authors, view caste-based discrimination as an inherently harmful and oppressive structure, we frame the breakdown in this paper as a 'value conflict'. We do so to focus on the procedural and relational challenges that arise when designers encounter deeply ingrained, and perhaps unreflexive, community practices that uphold such structures. Our goal is to analyze the negotiation process itself, rather than solely labeling the community's values as intentionally malicious.

\subsection{Reflexivity and Positionality}
\label{sec:positionality}

The first author is a female from the region where Seva is located. She speaks the local language and is familiar with the context. 
Growing up in the Backward Classes (BC), a middle tier in the caste hierarchy, she has an intimate awareness of caste dynamics. 
The organization is unaware of her caste, as it was never explicitly discussed. 
However, given that surnames often indicate caste, it is plausible that they could infer her background.
The second author is not from the region. While he does not understand the local language and nuanced differences in caste structures relevant to the context of the study, he grew up in a setting where similar caste hierarchies existed. As a person growing up in a Brahmin household, the dominant group in the caste hierarchy, he benefited, and continues to benefit, from the caste structure established in his region, including the privilege of not having to even think about caste growing up. 

We are outsiders. 
We write as outsiders, acknowledging our limited scope of influence within the context of our work. 
We know that the services provided by Seva are critical to a large population that does not have access to many other alternatives. 
As we write critically about our interactions with Seva, we also acknowledge that we are incapable of offering any of the essential services that Seva provides. 
We thus write as academics for an academic enterprise. 
We write this as critical social scientists with the hope that we can contribute the power afforded by our academic position to the broader and larger efforts being made for an equitable and just future free of the caste system. 

We both find caste structures to be problematic. We bring this value into the engagement, and the outcome we present in this paper is shaped by this value. 
We also note that the issue of caste is a social problem that requires social approaches. 
We believe that technology could play a role in enabling restorative justice and accountability, but a significant social approach is necessary to dismantle caste hierarchies that are deeply ingrained in South Asia, and, with the growing migration, outside of South Asia. 
We learn from Dr. Ambedkar and others \cite{ambedkar2022castes, criticalCasteSyllabus}, and echo their lessons that the existing structures need to be dismantled; gradual reform within the existing structure cannot lead to a just future.

\section{Literature Review}

\subsection{Caste and Design}

The caste system is a deeply entrenched social hierarchy in South Asia and is also practiced among South Asians beyond it. It is pervasive across various aspects of life \cite{zwick2018caste, conversation}; casteism manifests in both overt practices as well as subtle, everyday interactions. The system's pervasiveness is evident in its influence on social interactions, marriage patterns, employment opportunities, and access to education and healthcare \cite{vyas2022social}. Even in diaspora communities, caste-based prejudices continue \cite{yengde2015caste, zwick2018caste}.

Caste hierarchy finds its roots in Hindu scriptures and is enforced through Hindu ritualistic practices, including designating caste as a hereditary status, the commonplace practice of endogamy, and other overt and covert social segregation \cite{ambedkar2022castes, zwick2018caste, rajadesingan2019smart}. Some justify caste structures as an innocuous form of division of labor, ignoring social, material, and historical inequalities and discrimination that reproduce and sustain it \cite{ambedkar2002essential}. 
The pervasive institution of caste extends beyond its origins in Hindu society. 
It is also present in Muslim, Buddhist, Sikh, and Christian communities in South Asia \cite{conversation}. 
Moreover, caste is not confined to South Asia. 
As Murali Shanmugavelan puts it, ``The globalisation of caste is real'' \cite{criticalCasteSyllabus}.

To address the historical and systemic oppression of marginalized caste groups, the Indian government implements a policy of affirmative action known as reservations. 
This system aims to improve representation by reserving a percentage of seats in educational institutions, government employment, and welfare programs for individuals from constitutionally recognized categories: Scheduled Castes (SC), Scheduled Tribes (ST), and Other Backward Classes (OBC). 
While the exact quotas vary by state, access to these benefits typically requires individuals to provide official documentation (``caste certificate'') and declare their caste category on application forms.

Well beyond formal policy and documentation, caste continues to operate through a spectrum of normalized, everyday practices.
Scholars have documented how caste is often maintained through subtle actions: asking for surnames to infer caste background, assigning certain types of labor along caste lines, keeping separate utensils or water vessels, or avoiding physical contact under the pretense of ritual purity \cite{omvedt2019reinventing, yengde2019caste}. 
For instance, \citet{anjali2021watched} documents how caste surfaced in beauty work in Bangalore (India), where notions of hygiene and purity were used to assert caste superiority. 
These ``benign'' acts are performed without acknowledging casteist underpinnings, yet they reproduce deeply rooted social stratification.

Indeed, critical caste studies and emerging HCI literature highlight the persistence of caste in computing, challenging the notion of ``casteless'' modernity \cite{singh2024anti, vaghela2021birds,  vaghela2022interrupting, sp2022inheriting, kirasur2024understanding, omvedt2004untouchables}. 
\citet{vaghela2022interrupting} illuminate how caste practices continue to shape the technology industry in India and beyond. They show that the continued reliance on informal networks and referrals in hiring perpetuates caste-based exclusion. Beyond hiring practices, they highlight subtle ways in which caste-based biased practices are enacted in the workplace, such as in team dynamics, social interactions, and career support and progression \cite{vaghela2022interrupting}. 
Extending this, in a recent work, \citet{kirasur2024understanding} examine how casteist practices are perpetuated in online platforms both through overt hate speech and more insidious practices under the guise of humor or political critique.

Digital technologies are socially situated and thus both influence and are influenced by the social dynamics of the setting \cite{sp2022inheriting, toyama2015geek, nardi1996context}. 
Recent work illustrates this interlinked aspect of social and technical by examining data-driven systems and its role in perpetuating the marginalization of sanitation and domestic workers, who are predominantly from marginalized castes \cite{sp2022inheriting}. The work showcases how data-driven systems inherit and perpetuate discriminatory practices, and the opacity of the systems among workers exacerbates the power inequities, creating barriers to challenge the structural factors \cite{sp2022inheriting}. 
%Acknowledging unequal access to technology and inclusion both in the social and technical space, scholarship has called for designers to be more attuned to the socio-cultural contexts in which their design is situated. 

Scholarship in participatory design, critical caste theory, and social justice-oriented design emphasizes the importance of the social responsibility of researchers to ensure that they attend to the power dynamics of the settings \cite{winner2010whale, criticalCasteSyllabus, bratteteig2012disentangling, ogbonnaya2020critical}. 
In fact, there is a greater possibility that a design project may cause more harm than good if they do not attend to the social inequities that underlie the interactions in the setting \cite{masiero2022should}. 
Caste as a pervasive social practice perpetuates social inequities and harm; surfacing and attending to these issues become central responsibilities while designing technologies in these contexts. 
Recognizing the subtle-but-pervasive ways caste is practiced, such as in naming conventions, documentation practices, and institutional workflows, can help us understand how structural oppression is quietly encoded in and through technology.
In this paper, we share reflections on our attempt to attend to caste-based practices once they were uncovered, and the conflict and breakdown that followed.

\subsection{Conflict in Community-Based Design}

A growing body of CSCW and HCI scholarship has called for communities to be engaged in the design and research process, drawing from action research, participatory approaches, and design justice to democratize processes and challenge structural inequality \cite{hayes2011relationship, costanza2020design, duarte2018participatory, holeman2017co}.
The scholarship also highlights challenges and problematic elements in community-based design \cite{dombrowski2016social, cooper2022systematic, fox2014community, disalvo2010hci, jiang2022understanding, erete2023method}.
These include the extractive nature of research, which places an epistemic burden on the community \cite{pierre2021getting, bossen2010user}; a lack of engagement with the historic inequities present in the setting \cite{harrington2019deconstructing}; and assumptions of a homogeneous community that overlook local power dynamics \cite{bratteteig2016disentangling, saha2023benefits, le2015strangers}.
In all of the work, the degrees of community participation and the nature of participation vary.
Nonetheless, there is a wide consensus on the importance of building relationships with the community \cite{kotturi2024sustaining, le2015strangers, costanza2020design}.
This can be seen, for example, in the literature on building trust \cite{YeeWhite2015Goldilocks, clarke2021socio}, mutuality \cite{harrington2019deconstructing, liang2021embracing}, and long-term sustainability \cite{kotturi2024sustaining, kruger2021takes}.

What remains less explored, however, is how researchers should proceed when local values diverge sharply from their own, particularly when those values reinforce systemic oppression, such as caste-based discrimination.
As with other relationships, misalignments in values and conflicts between different community members --- and between community members and researchers --- are bound to occur.
Negotiating these tensions and conflicts when they emerge creates space for active involvement, influencing the collective vision for the future \cite{bratteteig2012disentangling, Falketal2022}.
Barring a few notable exceptions \cite{sawhney2020ecologies, hansson2016ting, jonas2022designing}, in CSCW and PD scholarship, conflicts between researchers and community partners are either under-reported or interpreted through framings that continue to prioritize relational harmony over structural justice \cite{gautam2024surfacing, park2022power}.

Along these lines, scholars building on Mouffe's ``agonistic struggle,'' argue for surfacing conflicts and differences between stakeholders \cite{bjorgvinsson2012agonistic, disalvo2015adversarial}. They encourage researchers to go beyond ``easy problem solving'' to confront ``the messiness'' inherent in the social context \cite{bjorgvinsson2012agonistic}. The goal here is to engage in ``transforming antagonism into agonism'' \cite{bjorgvinsson2012agonistic}, that is, not to resolve it \cite{gupta2021not} but rather enable pathways of mutual respect to act toward shared visions of the future \cite{bodker2018participatory, bjorgvinsson2012agonistic, disalvo2015adversarial}.
Democratic participation and the realization of just futures rest on such agonistic engagements and subsequent actions.
\citet{sawhney2020ecologies} introduce the notion of ``ecologies of contestation,'' encouraging design practitioners to anticipate and work with the messy terrain of conflict.
Similarly, \citet{umemoto2004walking}, in the context of participatory planning, cautions that even well-intentioned participatory processes may falter unless they engage with underlying epistemological frameworks and power structures that shape action and interpretation.
Their work suggests that agonism can be productive only when the designer is prepared to navigate power differences, engage with material and design constraints, and recognize their ethical position and complicity in the engagement \cite{sawhney2020ecologies}. 

However, agonism has limits, especially when values in conflict can lead to direct harm.
\citet{jonas2022designing} highlight this dilemma in their work with rural Appalachian communities, where researchers found themselves needing to navigate values they considered to be in opposition to their own (e.g., racism and sexism).
Through provocative reflections, they argue for continuing research with communities where researchers may have value misalignments and disagreements, but also emphasize that researchers ``should exercise their power to minimize or eliminate the negative impacts to indirect stakeholder groups who are marginalized by existing structures of power'' \cite[pp. 12]{jonas2022designing}.
This echoes \citet{bodker2018participatory}'s argument that participatory designers ought to take sides, particularly with those who are historically marginalized or harmed by the system.

However, when researchers take a stance, the relationship may break down entirely, ending any possibility of collective action toward a shared vision.
Our work presents one such scenario.
Our efforts to engage through negotiation and conflict failed.
CSCW and HCI scholarship offer little guidance on what researchers can do in such cases.

From one vantage point, this is an account of failure in research.
Following the ``turn to practice'' in HCI \cite{kuutti2014turn}, we reflect on the breakdown to reveal deeper structural entanglements and the need to rethink how researchers might strategically reposition themselves when structural harm is at stake.
We build towards pragmatic considerations which, while they may not prevent relationship breakdowns, could inform responsible and accountable practice in the face of systemic harm.

\section{Methodology}

Our partner organization, Seva\footnote{Pseudonym of the organization.}, was established in 2014. 
Since then, it has provided shelter to more than 3,000 survivors of domestic violence and mental health issues.
Central to Seva’s services are its shelter homes, which provide a physical sanctuary for women in need, along with 12 family centers (community centers) strategically located across a large city in South India.
The staff undertake wide-ranging, and often undocumented, tasks as part of their care. 
These can include helping survivors file police reports, accompanying them to hospitals, providing emergency hygiene and medical support, offering legal support, and providing counseling services. 
Not every client receives all forms of care; each ``case'' refers to the specific set of actions taken to address a client’s situation, which are recorded in a ``case file''.

Our understanding of the complexities surrounding the organization's efforts was limited. 
Thus, we engaged in an incremental study, where each phase of the study built on our findings from the previous phases. 
We planned this incremental approach to build trust through continuous engagement and to ground our design in the realities of the organization.

We commenced our study after obtaining IRB approval from our institution. 
The studies were conducted over two years. 
Due to resource constraints and visa complications, we were unable to travel, so the study was conducted remotely. 
This work was intended as the first step in a long-term collaboration with Seva.
We communicated with the staff members via Zoom and texted key personnel using WhatsApp, a messaging application widely used in India.
All staff members were comfortable with Zoom and WhatsApp.

\subsection{Research Approach}

In our research, we position ourselves as ``partners on the side'' \cite{gautam2024enhancing}, that is, we consider our primary role to be in supporting our partner organization in their efforts to support their clients. We sought to understand the organization's process and the staff's values and priorities. We did not engage directly with the organization's clients.

\begin{table}[]
\caption{List of participants in our initial inquiry. The subsequent phases of our work primarily involved Kavitha and Anita only. \vspace{-0.5em}}
\label{tab:participants}
\resizebox{\textwidth}{!}{%
\begin{tabular}{lll}
\hline
\textbf{Pseudoname} & \textbf{Position in Seva} & \textbf{Primary role in the organization}          \\ \hline
Kavitha &
  Owner &
  \begin{tabular}[c]{@{}l@{}}Establish and manage organizational processes;\\ Handle difficult client  circumstances\end{tabular} \\ \hline
Anita &
  Remote Coordinator &
  \begin{tabular}[c]{@{}l@{}}Handle coordination between counselors;\\ Manage client records and generate reports; \\ Manage communication with external entities\end{tabular} \\ \hline
Suma                & Program Coordinator       & Coordinate counselors through various programs     \\ \hline
Prudhvi             & Program Coordinator       & Coordinate counselors through various programs     \\ \hline
Aarya               & Counselor                 & On-the-ground staff interacting with their clients \\ \hline
Abhignya            & Counselor                 & On-the-ground staff interacting with their clients \\ \hline
Kamala              & Counselor                 & On-the-ground staff interacting with their clients \\ \hline
Sravani             & Counselor                 & On-the-ground staff interacting with their clients \\ \hline
Akanksha            & Counselor                 & On-the-ground staff interacting with their clients \\ \hline
Varun               & Legal Advisor             & Manage legal processes such as filing lawsuits     \\ \hline
\end{tabular}%
}
\end{table}

\subsubsection{Interview Phase}
We began with an exploratory initial inquiry to understand our partner organization's operations and staff members' values about their work. 
These interviews aimed to gather information about the internal processes of the organization, with the intent to examine the existing use of technology. 
The first author is a native of the region. She conducted the interviews in the shared native language. 
We believe that being a woman helped the first author in the interviews, as we felt that the interviewees felt more willing to open up to her. 
However, there were still challenges in communication arising from both the remote nature of the study and our positionality as privileged outsiders.

Each semi-structured interview was around one hour long. 
The interviews were conducted over two months. 
The decision not to interview further was made when we reached theoretical saturation, that is, we started hearing fewer new things as we continued our interviews. 
The overall theme of the questions revolved around the organization's services and the staff members' roles. 
We also discussed common issues they had noticed when working with their clients, and uncommon situations or cases\footnote{``Cases'' refers to individual clients \emph{and} the particular circumstance faced by the client that the staff is addressing. A single client can have multiple cases. These records of engagement, starting from intake, are stored in a ``case file''.} they had encountered.

In total, we interviewed ten staff members, all on Zoom. 
This included all the full-time permanent staff working in the main Seva Home. 
The interviewees included individuals working both at the main shelter and in the Family Centers. 
Predominantly, the staff members interviewed were women, with only two men (see Table \ref{tab:participants}). 
The two male interviewees, Suma and Varun, occupied more process-oriented roles within the organization, which may be attributed to the predominantly female clientele at Seva. 
One noteworthy aspect of the interview process was the reluctance of many staff members to participate in one-on-one interviews. 
As a result, most interviews were conducted with pairs, wherein both staff members shared their insights and responded to each other's thoughts. 
Anita suggested conducting interviews in pairs. 
This format mirrored the desire for collaborative work, where counselors and staff wanted opportunities to consult and learn from peers when handling cases.

\subsubsection{Prototyping and Technology Evaluation Phase}

We analyzed the interview data, which highlighted gaps in the organization's knowledge management system and limited structure in its processes (see Section \ref{sec:initialInquiry}). 
We recognize that these gaps stem from the resource constraints under which the organization operates.
Building on the findings, we designed a tool to support the organization's \emph{existing practices}, particularly in onboarding new clients and managing their records. 
This system would be made freely available to the organization, in contrast to the costly and complex alternative case management platforms that Seva had previously explored. 
During the interview phase, the organization shared its intake forms and seven case files. Our design was informed by these existing documents.  
We involved the organization's staff members, particularly Kavitha and Anita, at major milestones in this design process. 
We did this to incorporate the organization's preferences and requirements while ensuring we did not add to their already heavy workload.  
In those major milestones, we shared the designed artifacts with the staff and gathered feedback from them. 
This iterative design process involved creating wireframes and ultimately culminated in the development of a functional onboarding system.
The web interface was designed in English, as it was the preferred language for documentation and report generation.
We also discussed the possibility of incorporating the local language into the intake forms as a future enhancement.

\subsection{Data Collection and Analysis}

During our interviews, we took notes of our conversations and recorded the interviews. 
This resulted in audio recordings of around 15 hours. 
After the interviews, we began summarizing the transcripts at a sentence level. 
This formed our open coding process. 
We then moved the codes to a Miro board to conduct axial coding \cite{saldana2015coding}. 
In particular, we analyzed the relationships and hierarchies between the codes and how these relationships establish a better understanding of the organizational processes and priorities. 
Through multiple rounds of iteration involving both authors, we merged these codes to arrive at 20 higher-level codes. 
Some of the higher-level codes included ``issues in knowledge transfer protocol'', ``overworked staff'', ``technological resistance'', and ``strong reliance on other organizations''. 
While these codes help situate our broader findings, in this paper, we narrate our experience with some of the relevant themes (in Section \ref{sec:initialInquiry}) to highlight our design outcome (in Section \ref{sec:design}) and the subsequent conflict in value that arose (in Section \ref{sec:conflict}).

\section{Organizational Realities: Adaptive Care, Informality, and Discretion}
\label{sec:initialInquiry}

Our initial inquiry surfaced tensions between adaptive care practices and the absence of formalized mechanisms for coordination and institutional memory. 
The unstructured operations can be seen as Seva's strategies for meeting client needs within significant material and human constraints. 
At the same time, these adaptive approaches created space for discretionary practices that --- intentionally or not --- reproduced embedded social hierarchies. 
These themes shaped our design intervention and foreshadowed the conflict that subsequently emerged.

\subsection{Responsiveness as a Mode of Care}
Responsiveness was seen as part of the care that the organization offered.
A significant part of the organization's operations involves providing services to clients and coordinating with local support structures and institutions.
Providing this responsive care meant that many of the staff members had to be flexible. 
As Suma highlighted, ``\textit{My schedule changes every day. There are no specified daily routines. I visit the home daily.}'' 

This fluidity enabled personalized, compassionate care, but also made coordination difficult. 
Suma shared an example of such responsiveness in a seemingly simple case, ``\textit{If someone comes willingly without telling anyone, we inform the police, because their family may have put down a missing case.}''
The responsiveness extended beyond their work hours. 
Abhignya, a counselor, shared, ``\textit{I sometimes give my personal number to the clients in case they need to reach me}.''
Similarly, Prudhvi, a counselor, shared, ``\textit{If a client calls late at night and we get the call, I pick it up and try to offer her counseling through the call and tell her to come back in the morning so that we can better provide her support. But we first let her know that we are there for her/them}.'' 
However, not all counselors followed the same practices, reflecting differences in how care was delivered.

\subsection{Discretion and Variation in Decision-Making}

Without codified protocols, staff relied on personal judgment.
For example, in our discussion about how they decided what services to provide to clients in their first encounter, Suma shared, ``\textit{If they look fine, we don’t really take them to a health check-up. Only if there is a visible health problem will we take them to the hospital. Otherwise, we don’t take them.}''

More critically, these judgments were shaped by social cues that signaled client status. 
Staff explained that they adjusted their procedures based on who referred the client, ``[We]\textit{ don’t really inform the police about survivors who come from higher people},'' which when we inquired who constituted `higher people', it was clarified as ``\textit{media, NGOs }[other non-profits], \textit{and police stations.}'' 
Caste was not named explicitly, but these hierarchical distinctions mirror broader social stratification, including caste, class, and institutional affiliation, that structure deference and authority in the region. Over time, we came to see how distinctions were framed as pragmatic but pointed to variability arising from existing social hierarchies.

Opportunities for collective learning were limited.
The organization hosted monthly meetings among staff members, but staff members found that time was limited in sharing issues in cases. 
Staff used personal notebooks to document cases, often with little structure. 
As Sravani shared, ``\textit{We get a personal record book and we are able to record things about the cases or take notes about the individual clients as per our wish.}'' 
Some had formed a WhatsApp group for informal sharing.
However, this did not translate into a process for organizational memory or cross-learning. 

When asked about institutional guidance, Anita mentioned, ``\textit{Right now, there are no protocols that exist, and nothing is stored anywhere}.'' 
The lack of structure has several implications, including variations in services for clients, which, as we shall discuss later, became a central concern in our continued engagement.

\subsection{Overworked Staff and High Staff Turnover}

We could sense that the staff members were intrinsically motivated to help others. 
Aarya, one of the counselors, shared that ``\textit{Helper’s high is what motivates me}.'' 
Despite the intrinsic motivation, staff members reported being overworked. 
The wide-ranging responsibilities surrounding care and the absence of a formal organizational infrastructure had resulted in staff burden. 
We heard several accounts of staff members taking cases home and working well beyond standard hours.

In some cases, this resulted in corner-cutting to expedite onboarding or gain client compliance. 
One of the staff shared that she played ``\textit{good cop, bad cop}'' to get critical information from clients. 
A staff member shared an approach that involved intimidating the client to probe them for context regarding their situation, and another staff member recounted how she had ``\textit{forcefully removed her }[new client's] \textit{clothes}'' when the client needed to be cleaned but refused to undress. 

High turnover exacerbated these conditions.
Two of the staff members we interviewed, Abhingnya and Kamala, had started working in the organization less than three months ago. 
Staff members complained about the organization's lack of resources to hire full-time staff.
High turnover forced the organization to rely on interns for core services, including training and psychological assessment, which further created challenges in structuring their organizational processes.

\subsection{Reliance on Individuals, Not Systems}

%The lack of structured processes in the organization and limited opportunities for knowledge sharing had created a strong hierarchy among the existing staff members. 
Staff frequently deferred to Kavitha or Anita for significant actions and decisions.
In fact, in our conversation with Kavitha, she shared how when she is not around ``\textit{cases are pending so much. The workers }[staff members] \textit{don’t know what to do next.}''
Akansha echoed this, saying, ``\textit{We take the advice of Kavitha if it’s regarding a client that we’re not too sure about,}'' further adding, ``\textit{if it’s a client that’s a small issue, then we don’t take it up with her}.''
When we asked what constitutes a small case, she mentioned the subjective assessment of cases, ``\textit{issues that we are able to resolve, issues that don’t result in case, issues where the parties compromise}.'' 
This subjectivity was echoed in Aarya’s account of relying on Kavitha for guidance in counseling, ``\textit{... in family counseling, we ask Kavitha for further advice}.''

The dependence seemed to arise from structural factors rather than personal ones. For example, Anita's job included communication, report writing, initiating and participating in conversations, and creating monthly organizational reports by coordinating with staff members.
She stated that there was a ``\textit{lack of structure in the data being collected about the clients}.'' 
She shared a need to improve the organization's data collection processes, further highlighting their reliance on Kavitha's guidance to determine the type of data required for assessments, ``\textit{What data should I collect? What kind of data do I need for the report? I take input from Kavitha ... I would like to improve my data analysis skills, data collection methodology.}''

This interpersonal dependency, rather than a formal workflow, created bottlenecks and left little room for distributed decision-making or reflective organizational learning. 
This also meant that practices like recording caste could persist without institutional interrogation.
\\
% As noted above, staff members felt there were limited opportunities to share and build collective knowledge. 
% Clients' records were written down on paper and filed on shelves. 
% While Anita and a few staff members collected data from individual staff every month, it was primarily to create external reports and was not shared back with the staff members. 
% The staff members, including Anita, stated the limited opportunities to learn from  others and to grow in their jobs. 
% The monthly staff meetings, while helpful in supporting knowledge sharing, had no protocols in place to record and store the information obtained during these meetings for reviewing later.

These insights shaped the focus of our design intervention: to support organizational knowledge management and reduce staff burden on report generation without undermining the existing practices of care.
It is critical to note that the lack of resources, both human and material, influences each of the elements presented above. 
The lack of organizational process, extra individual efforts on the job, high staff turnover, and the reliance on key staff members all seem to point to the limited resources in the setting. 
We acknowledge and work within this context.

\section{From Initial Inquiry to Design: Web Application to Onboard and Record Services}
\label{sec:design}
We presented our preliminary findings to key staff members at Seva. 
They affirmed the challenges we noted, particularly the difficulty of information sharing, the burden of manual documentation, and the need for more systematic tracking of clients' journeys.
% They also proposed several ideas, including supporting clients in developing digital literacy, automating helpline call management, and facilitating staff members in following up with their clients. 
% %The centrality of technology reflects on our position as experts in technology. 
% We bring our technological expertise to the table; the direction that unfolded was influenced by our technological expertise. 
Indeed, one of the most prominent asks was ``\textit{If a woman comes into the organization, they take an application from the woman. This process is done manually. Can you create an application that automates this process?}''
Kavitha emphasized the need for digitization, ``\textit{How do we create a template or design or pathway where it is easy for us or anybody to understand problems we have and have a clear-cut pathway to recovery or rehabilitation.}''
Similarly, Anita asked if we could ``\textit{Obtain monthly reports of all the records they have typed and of the data they compile regarding the women and ... there is a report regarding the stats related to all the programs that Seva [anonymized] runs.}''

\begin{figure}
  \centering
  \includegraphics[width=1\linewidth]{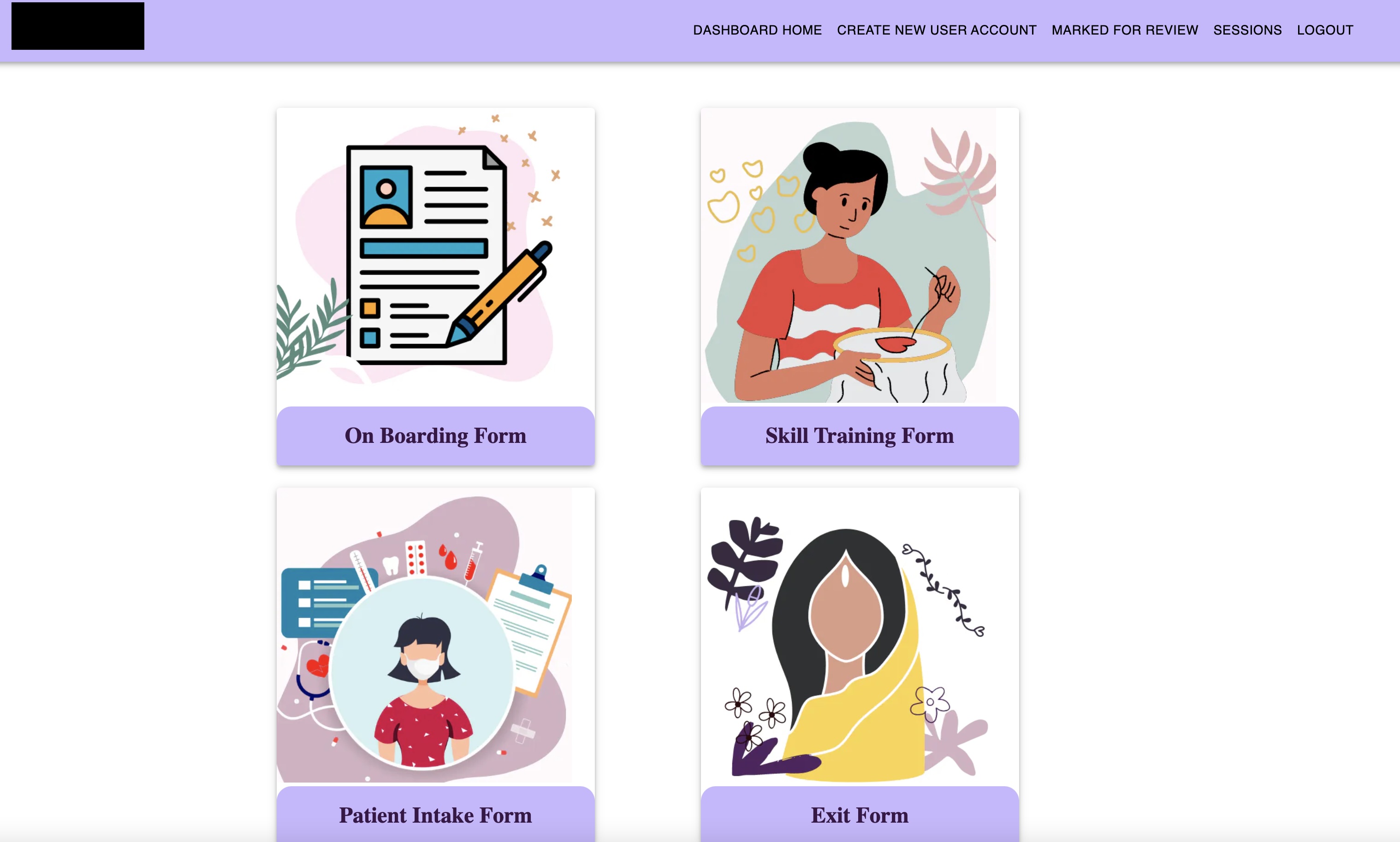}
  \caption{Our web application supported documenting four central forms that the organization used with their clients: the Onboarding form, Skill Training form, Patient Intake form, and Exit form.}
  \label{fig:homepage}
  \Description{This image shows the home page of our web application. It contains four cards with culturally appropriate icons. Clicking the cards will lead to different forms for the organization's staff members to fill out. The four cards are for the Onboarding form, Skill Training form, Patient Intake form, and Exit form.}
  \vspace{-0.5em}
\end{figure}

In response, we designed a web application to support their core processes that involved onboarding clients, documenting their journey within the organization (skills training, and healthcare services provided), and generating weekly and monthly reports shared with all the staff members. (Figure \ref{fig:homepage}). 
We sought to build upon existing forms and workflows to minimize disruption and promote comfort in using the tool.
Below, we briefly share our design considerations that involved aligning with the organization's existing practices and supporting knowledge sharing within the organization.

\subsection{Grounding Design in Existing Practices}

We wanted to ensure that the technology did not add overhead on already-overloaded staff members.  
To this end, the system was designed around paper forms previously shared with us.   
The organization had previously shared its forms and seven case files with us during the initial inquiry.  
This grounding allowed staff to transition smoothly to digital workflows without significant retraining or loss of continuity.

%We were careful about adhering to the existing paper-based forms when we created the forms on the web application.
The paper-based forms had questions to gather information about government IDs and resources such as Aadhar Card number (unique national identifier that contains biometric and geographic information), Permanent Account Number (PAN) card (unique ID for tax documents), and bank account details.
They recorded this information, as evidenced by the case files (client documents) shared with us, which had redacted markers. 
We inquired about the rationale for collecting this information, and the staff members stated that they use this information to support the clients in getting eligible government benefits. 
We collectively decided to make sensitive fields optional.
Further, to ensure safety, we stored the data in a HIPAA-compliant\footnote{The Health Insurance Portability and Accountability Act (HIPAA) is a US law to protect the privacy and security of individuals' health information. While HIPAA does not apply in India, where our work is based, it is widely regarded as the standard for personal data protection in technological infrastructures. In August 2023, the Indian parliament passed the Digital Personal Data Protection (DPDP) Act, which is largely equivalent to HIPAA. Technology infrastructures have now been developed to comply with the DPDP Act.} database.  

This was a context of mobile-first use \cite{donner2015after}.
All staff member had access to a desktop and/or a laptop, but some reported using only their mobile phones. 
They all used WhatsApp and emails, and were proficient with web navigation. 
% Other than Microsoft Excel and Word, none of them reported using any dedicated software. 
Considering this, we built a responsive web application that could be used both on computers and mobile phones.

\subsection{Supporting Knowledge Sharing}

Our design also sought to support the staff members in sharing information and learning from others. 
The organization documented the services on paper. 
Staff members reported struggling to find previous documents to understand the issues their clients had previously faced and the services they had received. 
Many of their clients were ``returners'', that is, individuals they had previously served. 
However, understanding their journey within the organization was challenging. 
To support this, we added a ``sessions'' feature that provided a running digital history of each client’s interactions so staff could track prior services and updates.
Similarly, we created a ``Mark for Review'' floating button on all four forms. 
This button enabled flagging specific entries that the counselors feel require special attention or further evaluation. 
These features were designed to shift the reliance away from individual's judgments towards institutional knowledge building and collective decision making.

\subsection{Evaluating the Technology with the Partner Organization}

We sought feedback from staff members at major milestones during the design process. 
This involved presenting a high-fidelity prototype of the web application to Anita and Kavitha. 

The prototype was, from our perspective, very well received.
Kavitha appreciated the feature to add sessions to each client's record, stating that ``\textit{it will be helpful}.'' 
They also mentioned the importance of adding an ``incomplete button'' on the form for administrators to mark if the counselors had not fully filled the form. 
When asked what they would use this for, they stated that they would like to use it to push counselors to fill in all the fields on the form, which included bank account details and Aadhar card numbers.  

These suggestions shaped our next iteration. 
While we hoped to support more reflective practices, we remained attentive to Seva’s priorities: easing data management, improving reporting, and preserving continuity amid turnover.
As we describe next, it was in this context of ongoing collaboration that a deeper value conflict around the use of caste fields in the forms surfaced.

\section{Issue of Contention: Caste Field}
\label{sec:conflict}

The staff members had expressed enthusiasm for using the web application when we showed them the prototype.
We gathered this reaction as a positive outcome for our design research. 
Building on this enthusiasm, we created user manuals for both administrators and counselors. 
We demonstrated the web application, highlighting the changes we had made based on the previous round of feedback and providing guidance on the different affordances. 
We deployed the web application and made it available to the staff members at Seva. 

However, one critical difference had remained unaddressed during the prototyping phase: the omission of the caste field from the digital forms. 
On the organization's paper-based forms and case files, we noted that the staff members had recorded the caste of all their clients.
These caste fields were open-ended text entries that specified a client’s detailed caste identity rather than using broader constitutional categories like SC, ST, or OBC.
%\footnote{In India, the government recognizes three caste categories for affirmative action: Scheduled Castes (SCs), historically marginalized and often referred to as Dalits; Scheduled Tribes (STs), comprising many indigenous groups; and Other Backward Classes (OBCs), which include socially and economically disadvantaged castes.}. 
While these granular markers are often used informally in speech, recording them in institutional systems can reify caste hierarchies and expose clients to stigma. They are rarely necessary for eligibility documentation, which typically requires the broader categories.
Given the wide but often subtle prevalence of casteist discrimination across South Asia and beyond, with very limited accountability and repercussions (e.g., \cite{vaghela2022interrupting, nytCasteUS, ambedkar2022castes}), we saw it as problematic to codify castes into digital forms.

This decision was rooted in our values as researchers (as noted in Section \ref{sec:positionality}), and in retrospective readings of earlier interviews, where discretion in services based on client background was evident (Section \ref{sec:initialInquiry}).
For example, referrals from ``higher people'' led to differences in the treatment of the clients. 
Even though it never emerged in our interviews, caste discrimination could have influenced some of the decisions regarding the services.
We feared that digitization could entrench and amplify structural harm.
So, in our digital forms in the web application, we did not include caste fields.

Until this point in the engagement, we had not discussed our decision to omit the caste field with the staff.
After the demonstration, building on the positive feedback we had gathered so far, we began inquiring about the caste field.
We did this to understand why the text-based detailed caste marker was important and build on it to engage in reflective conversation and collective decision-making.

\subsection{First Iteration: Caste as Pragmatic Necessity}

When we presented the prototype, the staff members requested that we add the caste fields to the form.
Kavitha responded, ``\textit{caste is very important, please add it in the forms}.''
Anita supported this idea, emphasizing that identifying caste can provide critical references for data analysis to show the organization's impact.

This was a point of conflict that we felt was critical to engage with so that we and our partner organization could move forward together with a shared anti-caste vision.
We inquired why they felt caste was important to include in the forms, especially given how including it in the digital forms would formalize the practice of marking caste within the organization.
Kavitha stated that ``\textit{caste-based entitlements are big things in India},'' and thus knowing a client's caste would help them in obtaining those services. 
Anita insisted that they use castes to remember or refer to clients, elaborating with an example, ``\textit{If there are two people whose name is Padma, we will use caste as a means to distinguish them ... Swakula Padma.}''
``Swakula'' is an explicit caste marker. 
The use of caste qualifiers to differentiate people, instead of, say, the more common practice of place (e.g., Padma from OMR and Padma from Church Street), signifies the centrality of caste in the organization's operations. 
It also reflected a distance between the staff using such markers and the clients being marked by them.

We began connecting this qualifier with the previous accounts of differences in services provided to the client, even though caste was not explicitly named as a factor.
When we inquired why none of the staff members mentioned the importance of caste earlier in our interviews, Anita stated, ``\textit{they are too shy and they don’t know, but caste is ingrained in us,}'' further elaborating the privilege of not engaging with caste, ``\textit{We are from upper caste so we come from a place of privilege, but not all others [clients] are extended that.}'' 

This moment clarified the misalignment. 
While we were focused on caste-based structures, the staff viewed caste data as a pragmatic and essential input for service delivery, particularly for accessing state-sponsored welfare programs. 
We proposed a set of radio buttons listing the constitutional categories: Scheduled Castes (SC), Scheduled Tribes (ST), and Other Backward Classes (OBC), rather than a free-text field.
We hoped this would \emph{satisfice} \cite{simon1997models}, preserving the functionality needed for benefit eligibility while reducing specificity that might perpetuate discrimination. 
We reasoned that there would be no room for further discussion and change if we did not move forward with this impasse. 
We assumed this compromise might open space for future discussion and reflection. 
We were wrong.

\subsection{Second Iteration: Institutionalization of Caste Through Practice}

We incorporated the radio button option into the revised prototype and presented it again. This time, the response was more forceful. 
One staff member stated, ``\textit{See in the country} [India] \textit{all the benefits are based on the caste. Our society itself is built on the caste, so it is an important social marker. If we ignore that basic, important social marker, we are actually not understanding one ... we don't actually understand their social background ... and we cannot get the entitlements because all entitlements are caste-based ... everything is caste-based in the country.}''
They further emphasized the centrality of caste in India ``\textit{it's our society, whether we like it or not, is a caste-based society.}''
Staff members also shared different fee waivers provided to different caste groups as justification for recording the castes of individuals.
This led a staff member to comment that knowing the caste of an individual is more important than knowing their age.
We noted the implicit bias centered on the claims of supporting women empowerment when a staff member shared, ``\textit{And also if we don't get ... demographic where a woman from SC or ST have more domestic violence or more problems if we don't get that the idea we can't you know that data will be very useful for policymakers. For example, more interventions need to be done for women empowerment in ST areas or SC areas is something policymakers and researchers all need that information. That is the reason that we are asking for that to be included.}'' 

These comments highlight the crux of the value conflict. 
First, it perpetuated negative stereotypes of marginalized caste groups. 
Second, it highlighted the distance between the staff and the clients they were serving. 
Those from dominant castes often have the privilege of viewing caste as a benign social marker, failing to recognize the subtle but pervasive harm it enacts, the privilege that a staff member had even alluded to earlier.
More importantly, even in the example they gave, recording affirmative action categories (e.g., SC/ST/OBC) through a radio button would suffice.
We reiterated that access to affirmative action policies could be met with records of broader caste categories, without needing the granular, free-text field.
However, they objected. 
This objection did not stem from logistical concerns. It was an objection to our questioning of a broader cultural orientation in which they saw caste as a natural and essential aspect of understanding clients.
From here on, we could not arrive at a collective recognition that collecting and recording caste might itself reproduce the hierarchies they sought to navigate. 
% Instead, caste was framed as neutral data, a demographic variable critical to institutional operations.
% We sensed resistance not to our particular decision on the interface, but to our broader stance of caste as a structure to be challenged rather than accepted as a bureaucratic necessity.

Noting the lack of room for discussion and negotiation, we presented the radio-button-based caste field. 
We requested them to try it in their everyday operations and share feedback.
The staff agreed.
However, they never used it.
We followed up with multiple messages and emails, but we never heard back from them. 
Much later, Anita, in a WhatsApp message, told us that the organizational priorities had shifted, which meant they would not be using the web application. 
We interpret this as a quiet breakdown, a gradual unraveling marked by silence and distance.

\section{Discussion}

Our two-year engagement with Seva, a non-profit organization in southern India that supports women seeking support after domestic violence and mental health crises, sheds light on some of the complexities and challenges of community-based design. 
%We highlight critical issues that arise when attempting when working across cultural and geographical distances.
Through the initial inquiry, we found that some of the organizational issues were amenable to design interventions.
In particular, technology could support ongoing efforts by facilitating client onboarding and organizational knowledge management. 
The relatively low cost and sufficiently abstracted web development ecosystem enables us to make technology a feasible solution for significant organizational challenges.
This reflects the positive outcome of human-centered design methodologies central to community-based design work.

In our case, our perception of the social repercussions of technological development and deployment created conflict that culminated in the breakdown of our collaborative relationship.
A system that prominently records caste could inadvertently contribute to the persistence of caste-based discrimination, even if that is not the intent of the stakeholders.
While caste information might initially be used for benefit allocation, future uses could include discriminatory practices in employment, housing, or other social services.
Moreover, the long-term consequences of recording such sensitive social categories are often unknown.

\subsection{Challenge in Balancing Tensions on the Ground}

We find ourselves torn.
Seva is an organization that provides services in a context where few are available. 
There are several intersecting injustices: everyday caste discrimination in the setting and the reality of gender-based violence faced by Seva's clients, who do not have access to many resources for support. 
Seva's services and practices actively address these dynamic forces. 
Moreover, as our findings show, the staff members cared for their clients. 
We think our responsibility is to be partners on the side, bringing institutional power and socio-technical knowledge to support the organization in its ongoing work. 
By taking a critical stance, we did not offer any direct support to our partner organization.

While we recognize casteism as inherently harmful, we frame this as a value conflict to emphasize the relational and procedural challenges of confronting deeply embedded practices, some of which may not be easy to discern.
Some of the practices we observed arose from resource constraints.
The ``corner-cutting'' practices, such as collecting bank account information or playing ``good-cop, bad cop'' roles, can be interpreted as practical responses to low resources and challenges in the setting. 
In the face of a large number of cases, such strategies may be required.
While we do not condone such practices and wish to see more resources and support available so that these practices would not continue, balancing the tensions on the ground entailed acknowledging and seeking to design within those realities.

In the case of eliciting and recording specific castes, however, we felt we needed to draw a line.
From our standpoint, including a free-text caste field would have codified caste within the organization's new workflow and risked reifying a deeply oppressive social structure. 
It felt more than an issue of data representation as it could expose clients, many of whom are from marginalized communities, to further stigma and harm. 
These clients, as indirect stakeholders, had no voice in shaping the technology, but their lives would be shaped by its use. 
As researchers and designers, we are responsible to the indirect stakeholders affected by the technologies we help create. 
Indeed, researchers ought to ``take sides'' \cite{bodker2018participatory}, and, to arrive at just futures, this involves taking sides with those who are most marginalized by existing structures. 
% by making efforts to encode liberatory values \cite{costanza2018design, collective1977black} that counter oppressive structures such as the caste system.

For us, there was no possibility of balancing the tension in values. 
We began from a position open to negotiation \cite{gronvall2016negotiation}, but we remained firm on our position and instead sought to bring the partner organization to our position.
That did not happen.

\subsection{Technology-in-Practice and the Reproduction of Caste}

Importantly, the staff members did not explicitly express casteist beliefs.
They cared about their clients.
Indeed, as our early findings show, many of them went beyond their means to support those they served.
Yet, casteist practices were present, not out of overt ideology, but through the enactment of existing practices. 
The staff were navigating institutional constraints, including state-mandated caste-based documentation for welfare access.
Their request reflected institutional realities. 
At the same time,  their insistence on preserving detailed caste identifiers reflects the deeper cultural centrality of caste, rather than a bureaucratic necessity. 
The social distance between the staff and their clients enabled this uncritical acceptance of caste as a routine part of their practice.
%the privilege of not noticing caste is afforded to those from dominant caste groups, a point the staff acknowledged.

The use of caste markers like ``Swakula'' in both formal documentation and informal discourse revealed how caste had become institutionalized within the organization.
The free-text caste field, when used routinely, became a normalized element of Seva's intake process.
It shaped and was shaped by existing social structures, forming part of what \citet{orlikowski1992duality} describes as the structural properties of organizations.

We argue that had we included an open-ended caste field, the technology (the digital form) would have reified and legitimized that practice despite our reservations.
Technology carries an illusion of neutrality \cite{ames2015charismatic, feenberg2010ten}. 
But technology is not neutral; it is constituted by and constitutive of the social practices around it.

Technology does not merely reflect social values. 
It can codify political decisions into seemingly immutable forms \cite{benjamin2019race, winner2010whale, noble2018algorithms, o2017weapons}. 
Institutionalized technologies embody social norms and rules, and through repeated use, enact and reinforce those norms as structural reality \cite{orlikowski1999technologies}.
In this way, technology can amplify existing inequities rather than correct them, particularly when practices are taken for granted as neutral or beneficial \cite{barocas2017engaging, toyama2015geek, sp2022inheriting}.

Institutional harm can emerge without explicit values. 
It can stem from seemingly routine, neutral actions.  
The staff may not have recognized that caste was being enacted through the organizational practices. 
From our perspective, however, the patterns became visible, and it was this difference in recognition and response that contributed to the conflict.
We echo \citet{frauenberger2019entanglement} in arguing that, as both technology designers and human beings in a world with deeply embedded structural inequities, we must engage with the entanglement between technologies, social systems, and the everyday practices that enact and reinforce those systems.
Separating the technological from the social is a grave fallacy. %, especially in contexts where sociotechnical systems quietly sustain and legitimize longstanding hierarchies.

\subsection{Limitations}

Before drawing out the implications of our work, we discuss the limitations.
Our study has several limitations. 
A remote study does not afford relationship building, which may have allowed us more opportunities to notice differences in our values and perhaps even support taking multiple small steps in conversations when contention arose. A remote study meant that each interaction carried greater weight.  
The remote interactions may have hastened the breakdown and rendered it harder to repair. 
However, we also contend that the phenomenon of interest in this paper --- value conflict among a key community stakeholder and researchers --- would have arisen even in an in-person study. 

Similarly, our study lacks concrete evidence of the relationship breakdown. As with personal relationships, a breakdown entails less communication, which, for academic reporting, is challenging. In fact, a cursory reading of Anita's remark about the shift in priority would suggest that there was no such breakdown. We rely on our inference, particularly of what we feel and sense. We report the drastic shift in the enthusiasm for the web application and the change in communications after the conflict arose. We base the presentation of our findings on what we felt from the conversations before and after the conflict. We acknowledge this as a limitation. We also mention it here to make epistemological trouble \cite{harrison2011making} and call the CSCW and HCI community to expand what counts as valid ways of knowing beyond the logical and textual. This is particularly important when rupture, silence, refusal, or withdrawal are part of the findings. 

Mutuality and reciprocity are cornerstones of community-based design \cite{dreessen2020towards, cooper2022systematic}. But our work is limited on this front. 
We are reporting on disagreements between us as researchers/authors and the partner organization who do not have an equal voice in this paper.
More critically, the organization did not get any meaningful outcome from their participation in our study. 
We had initially envisioned a long-term collaboration, but the breakdown disrupted that path.
In May 2024, after it became clear that the relationship might not be repairable, we created a slide deck distilling the findings reported in this paper. 
We shared this with the staff in the organization through email. 
As of this writing, we have not heard from them.

\subsection{Implications for Community-Based Design}

From our standpoint, the engagement was a failure because we were unable to illuminate broader possibilities for and with our partner organization.
We re-imagine what we may have done differently. 
The experience also challenges our expectations of what community-based design can achieve in contexts where structural issues inform and are reinforced by local practices.

Building on our empirical findings and reflections, we articulate four modes of engagement for navigating value misalignments in community-based design. 
These modes of engagement would have helped us better calibrate to the relational, institutional, and political conditions of the setting. 
Each mode reflects different possibilities for acting with care, before, during, and after the moment of conflict.
In all of these, we need to ensure that our value stances are shared --- an anti-caste approach, in our case --- for without doing so, there may not be any pathway toward justice \cite{costanza2020design, collective1977black}.

\subsubsection{Mode 0: Cultivating Trust and Relational Conditions}
\label{sec:mode0}

Before critical reflection (Mode 1) or distancing (Mode 2) can unfold, community-based engagements require a foundation of trust.
Trust is not a singular state \cite{mollering2001nature} but a set of evolving conditions that shape what forms of actions, reactions, and critiques are possible and acceptable. 
In settings where structural inequities are present, it is especially important to recognize that trust may be uneven and dynamic, tied to power and positionality, and susceptible to quick erosion.

The remote nature of our study limited our ability to build deep trust. 
We lacked the informal interactions, shared meals, and rhythms of co-presence that often foster mutual understanding and allow for dissent without defensiveness. 
This did not cause the value conflict we describe, but it likely shaped how it unfolded, making it harder to probe uncomfortable topics, or detect resistance early.

We propose that community-based researchers consider trust as an ongoing variable to be attended throughout the engagement \cite{corbett2018going, harding2015hci}. 
Higher levels of trust may enable direct reflection and challenge, whereas lower levels may require more scaffolded or indirect methods. 
Trust is not built just by being present.
It also involves transparency about one’s values and consistency in commitments \cite{corbett2018going}. 
Building trust is thus \emph{not} a pre-requisite activity but an ongoing mode of engagement that strengthens all others.

\subsubsection{Mode 1: Reflection and Situated Alignment}
\label{sec:mode1}

We believe that promoting critical reflection with staff could have helped surface implicit values and institutionalized practices, opening pathways for alignment building upon the shared values.  
Drawing on scholarship in speculative design and practice theory \cite{kuutti2014turn, wong2017eliciting, dillahunt2023eliciting}, we offer two strategies that can support reflection on embedded institutional practices.

\paragraph{Speculative Design for Institutionalized Practices:} 
Speculative design is widely used in CSCW and HCI scholarship. 
It has been used to deepen participant engagement in research and elicit values (e.g., \cite{wyche2021benefits}).
It also enables people to construct alternative worldviews \cite{wong2017eliciting}.  
We propose to use it for reflecting on current practices. 
Conflicts are often rooted in epistemological differences \cite{umemoto2004walking}.
Seva’s caste-based workflows may not necessarily be ideological assertions, but institutional logics emerging from a bureaucratic epistemology of service provision (e.g., state reporting requirements) that became institutionalized. 
As we learned from our later conversation, staff members hesitate to speak about caste-based practices, many of which are emergent and implicit.
By creating provocations that exaggerate current workflows (e.g., imagining future scenarios where caste data is leaked or misused), we might have supported staff in stepping outside their current practices and exploring alternatives.
In this respect, we position speculative design as a research method to be used as part of the work before \emph{the} work (of design).

\paragraph{Dynamic Power Mapping:} 
We noted the hierarchy within the organization, and we moved around without questioning that hierarchy. 
More critically, since we were designing for the organization, we did not explicitly raise questions on the power differences between the staff members and their clients. 
Indeed, in such structures of dependency, where clients are heavily reliant on the organization, a direct confrontation is untenable and could even be harmful \cite{gautam2022empowering}.
Looking back, incorporating power mapping to collectively reflect on the power structures between different staff members and their clients could have been beneficial \cite{schiffer2007power, hagan1997power}. 
Power maps enable the identification of key stakeholders, their relationships, and the loci of power. They are widely used in community development and organizing efforts (e.g., \cite{ucs_powermapping_2018, nea_powermapping101_2023}).
However, rather than creating power maps as static documents, we argue that they should enable dynamic representations and reflections of the relationships and influence within the organization. 
This could be realized, for example, using a shared digital power map, updated iteratively, that makes visible the informal networks, gatekeeping roles, and technology-mediated influence (e.g., WhatsApp group admins) that shape decision-making. 
The use of a power map is to surface where decision-making is concentrated and what practices are open to change.

These reflections on institutionalized practices and power structures are aimed at arriving at possible pathways for alignment. 
This necessitates that we, as researchers, reflect on our own values and identify what we prioritize in the engagement. 
As \citet{jonas2022designing} posit, we should focus on shared values during the engagement. 
This involves the ``exclusion of values which are opposed by \emph{any} stakeholders'' \cite[pp. 8]{jonas2022designing} (emphasis in original text). In this context, researchers are stakeholders.
However, as we experienced, this may sometimes rupture the relationship.
Thus, we advocate for a different approach, one that brings to the fore values that are important to all stakeholders through the reflective exercise, and then, if common ground seems untenable, shift to collaborating from a distance.

At the same time, we advocate for care in facilitating these exercises. 
In an under-resourced environment like Seva, where staff are already overworked, such exercises may be perceived as an abstract academic luxury, disconnected from their urgent, practical needs (e.g., \cite{harrington2019deconstructing}). 
%This could strain the relationship and reinforce the idea of researchers as outsiders who do not understand the ground realities. 
Moreover, careful facilitation has to be matched with actionable steps for change so as not to reinforce existing power dynamics or promote a false sense of change.

\subsubsection{Mode 2: Collaboration from a Distance}
\label{sec:mode2}

When fundamental values diverge, designers must be able to step back without entirely stepping away. 
It is possible to \textit{collaborate from a distance} where we acknowledge differences, but where mutual respect and transparency support continued-but-limited engagement.
Collaboration from a distance can help build reciprocity in the relationship, where we would continue to remain available as supportive but critical partners. 
This requires opening space for relational plurality, where we could collaborate on shared goals without full agreement on values. 

Designers must consider forms of engagement that do not rely on persuasion or consensus. 
Our refusal to implement the caste field was one such case, but it was not accompanied by a structured approach for what to do next. 
Below, we reflect on two strategies structured to keep the possibility of mutual engagement open. 
It requires openness to the fact that the relationship may break down despite the effort.

\paragraph{Distance through Role Reframing}
We believe that all stakeholders, including researchers, need to articulate their values. 
Mutual engagement requires this, as we evolve together in the influence of the other. Such conversations require surfacing non-negotiables on both sides without assuming full alignment. 
Once the values are shared and known to all stakeholders, we establish boundaries of our engagement. 
This practice parallels approaches in collaborative governance and agonistic negotiation \cite{gronvall2016negotiation, bodker2018participatory}.

In our case, we do not regret taking a stance against building the caste fields.
However, we could have articulated our boundary of involvement more explicitly, that is, where we could support the organization in the ongoing efforts (e.g., grant writing or digital literacy), and where we could not (e.g., building tools that codify caste). 
We remained focused on the onboarding tool (Section \ref{sec:design}), whereas we could have sought to open new threads of partial presence with Seva.

\paragraph{Asset-Based Appreciative Inquiry} 

In hindsight, refusing to implement the free-text caste field could have come across as a refusal of broader organizational practices.
This was not our intent, but it points to how value conflicts can be perceived as a moral indictment.
Instead, we could have engaged as assets-based designers who focus on initiating collaboration by identifying and amplifying the resources and strengths already available in the setting. 
Assets-based design does not ignore the needs or systemic issues but rather first seeks to collectively identify and build upon available assets to address them \cite{wong2021reflections, pei2019we}.
In our case, this could have involved drawing out the staff members' caring practices and the organization’s gender violence response protocols as foundations for re-imagining other forms of onboarding practices. 
The shift from deficit to assets can help preserve dignity and trust during moments of ethical divergence.
Further, building on existing assets could open up room for changing practices that are based on the organization's strengths and capabilities.

\subsubsection{Mode 3: Exiting with Care}
\label{sec:mode3}

Collaboration from a distance may not always be viable. 
When continued engagement, even when limited, may risk legitimizing or being complicit in harm, designers may need to consider exiting. 
Academic collaboration must not come at the cost of reinforcing structures that can harm.
Shifting from distancing to exiting should be driven by a reflective assessment of whether the engagement still holds space for mutual influence and learning. 
If no such space remains, our responsibility entails exiting deliberately and with care. 

In our case, after we refused to implement the caste field and the organization quietly disengaged. 
We chose not to push further beyond the occasional check.
Much later, we shared our findings in an accessible format, outlining both the rationale for our design decisions and the assets we saw in their work.
While this was a step toward care, it was not accompanied by a more accountable conversation or closure.

Just as we enter and build relationships with care, we must also exit with care.
Doing so may involve sharing documentation and other resources, naming harms, or simply acknowledging the relationship’s value. 
Even when exiting, we must hold ourselves accountable by justifying why and how we are exiting.
\\

These modes of engagement form a repertoire for community-based design to draw upon in response to evolving conditions, such as shifting trust, emerging tensions, or visible harms to direct or indirect stakeholders.
They are interwoven strategies that may co-exist. 
Decisions about which mode to adopt and when require ongoing reflection on the relationship dynamics, institutional constraints, and our value commitments.
Our work must remain grounded in our values, even when they lead us into conflict.

\section{Conclusion}
This paper presents a case study of a relationship breakdown during community-based design, shaped by tensions around the institutionalized practice of caste. 
We discuss how we perceived the often subtle caste was embedded in institutional routine practices, which we realized later in the research process. 
As we engaged in design, we began anticipating the potential harm that our designed technology could inflict by perpetuating and codifying existing social inequities. 
The conflict and the subsequent breakdown of our collaborative relationship highlight the ethical dilemmas that community-based designers face, and the social responsibility that designers need to undertake. 
Our journey prompts the necessity to attend to institutionalized practices --- the activities that are enacted by and enact the social structures --- early in the engagement. 
We advocate the need to build a rich repertoire of engagement, where reflection, strategic distancing, and values alignment are foregrounded to challenge injustice while attending to local realities. 
Designing with communities must include space for disagreement, discomfort, reconfiguration, and refusal, and not just consensus.

\begin{acks}
We are grateful to our partner organization for working with us on this research project. This academic outcome would not have been possible without their collaboration. We are also deeply grateful to the anonymous reviewers at CSCW and ICTD  for their constructive feedback. We were uncertain how to report failure, and their suggestions and guidance were invaluable in shaping this paper in its current form.
\end{acks}

\bibliographystyle{ACM-Reference-Format}
\bibliography{sample-base}

\end{document}